\documentclass[preprint,aps,prc,amsmath,amssymb,
runinaddress,
showpacs]{revtex4-1}
\usepackage{amsthm}
\usepackage{graphicx}
\usepackage{dcolumn}
\usepackage{bm}
\usepackage{lineno}

\begin{document}

\title {Extracting temperature and transverse flow by fitting transverse mass spectra and HBT radii together}

\author{Ronghua He}
\email{ronghuahe2007@163.com}
\affiliation{Department of Physics, Harbin Institute of Technology, Harbin 150001, People's Republic of China}
\author{Jing Qian}
\affiliation{Department of Physics, Harbin Institute of Technology, Harbin 150001, People's Republic of China}
\author{Jianyi Chen}
\affiliation{Department of Physics, Harbin Institute of Technology, Harbin 150001, People's Republic of China}
\author{Qingxin Wu}
\affiliation{Department of Physics, Harbin Institute of Technology, Harbin 150001, People's Republic of China}
\author{Lei Huo}
\affiliation{Department of Physics, Harbin Institute of Technology, Harbin 150001, People's Republic of China}

\date{\today}

\begin{abstract}
Single particle transverse mass spectra and HBT radii of identical pion and identical kaon are analyzed with a blast-wave parametrization under the assumptions local thermal equilibrium and transverse expansion. Under the assumptions, temperature parameter $T$ and transverse expansion rapidity $\rho$ are sensitive to the shapes of transverse mass $m_\text T$ spectrum and HBT radius $R_\text{s}(K_\text T)$. Negative and positive correlations between $T$ and $\rho$ are observed by fitting $m_\text{T}$ spectrum and HBT radius $R_\text s (K_\text T)$, respectively. For a Monte Carlo simulation using the blast-wave function, $T$ and $\rho$ are extracted by fitting $m_T$ spectra and HBT radii together utilizing a combined optimization function $\chi^2$. With this method, $T$ and $\rho$ of the Monte Carlo sources can be extracted. Using this method for A Multi-Phase Transport model (AMPT) at RHIC energy, the differences of $T$ and $\rho$ between pion and kaon are observed obviously, and the tendencies of $T$ and $\rho$ vs collision energy $\sqrt{s_\textmd{NN}}$ are similar with the results extracted directly from the AMPT model.
\end{abstract}

\maketitle

\section{Introduction\label{sec:introduction}}
Experiments of heavy-ion collisions aim to find and study quark gluon plasma (QGP). The created hot and dense nuclear matter cools down and transforms to hadrons. The produced hadrons interact with each other until freezing out at the final stage. The space-time information can not be measured directly, and some indirect observations are utilized to get the dynamical information, such as transverse mass ($m_\text T$) spectra \cite{mts, mts2, mts3, mts4, mts_exp, mts_exp62GeV, mts_v2} and two-particle correlation functions \cite{hbt_sum, hbt_sum2, hbt_sum3, hbt_sum4, rs2_rms, rs2}.

Transverse mass spectra of freeze-out hadrons can be influenced by many factors, such as resonance, jet, temperature and transverse flow. Resonance is considered to influence transverse mass spectra at low $p_\text T$ \cite{hbt_resonance}. Jet is consider to be dominate of $m_\text T$ spectra when transverse momentum $p_\text T$ is larger than 2 GeV. The dominate factors of the intermediate transverse mass spectra are considered to be temperature $T$ and transverse flow $\rho$ \cite{exp_expansion, exp_expansion_hbt_mts, exp_expansion_hbt_mts2, exp_soft_mts_hbt, mts2013JPG}. But it is difficult to extract temperature and transverse flow from transverse mass spectra of one kind of particles, because the transverse mass spectra with low temperature and high transverse flow rapidity are similar to the spectra with high temperature and low transverse flow rapidity \cite{LQShan} (described in detail in Sec.~\ref{sec:method}). In the past, spectra of different kinds of hadrons were fitted together to extract $T$ and $\rho$ under an assumption that $T$ and $\rho$ are same for different hadrons, but this assumption can not be proved directly.

On the other hand, identical particles correlations show another way for extracting temperature and transverse flow. It was observed in experiments that HBT radii $R_\text s$ decrease with averages transverse momentum $K_\text T$ of pairs of identical pions or identical kaons, and this phenomenon is considered to be dominated by transverse flow \cite{exp_expansion, exp_expansion_hbt_mts, exp_expansion_hbt_mts2, exp_soft_mts_hbt, rs2_kt, hbt_pion_exp}. Opposite to the phenomenon of transverse mass spectra, temperature and transverse flow have a positive correlation for HBT radius $R_\text s(K_\text T)$ \cite{LQShan}. In other words, the shapes of $R_\text s(K_\text T)$ with low temperature and low transverse flow rapidity are similar to those with high temperature and high transverse flow rapidity, so transverse mass spectra and HBT radii can be utilized together to extract temperature $T$ and transverse flow rapidity $\rho$ \cite{exp_expansion_hbt_mts, exp_expansion_hbt_mts2, LQShan, model_hbt_mts}.

For identical pion, $m_\text T$ spectra and HBT radii $R_\text s (K_\text T)$ was fitted respectively to get the equi-error contours on $T$-$\rho$ plane, and the crossing area of two kinds of contours are considered to be results of optimization \cite{BlastWaveMtsRs2, BlastWaveMtsRs2_2, LQShan}. In this paper, for both identical pion and identical kaon \cite{hbt_kaon, hbt_kaon2, hbt_kaon3, hbt_kaon4, hbt_kaon4a, hbt_kaon5, hbt_kaon6}, we try to use a combined optimization function $\chi^2$ including both $m_\text T$ spectra and HBT radii $R_\text s(K_\text T)$ to extract $T$ and $\rho$, and the reason why these two kinds of data can be fitted together are clarified by utilizing maximum likelihood method.

This article is organized as follows. Under the assumptions of local thermal equilibrium and transverse expansion, blast-wave function is utilized to deduce the expressions of the $m_\text T$ spectra and HBT radius $R_\text s(K_\text T)$, and the correlations between temperature $T$ and transverse flow rapidity $\rho$ will be discussed in Sec.~\ref{sec:method}. In Sec.~\ref{sec:mc}, for pion and kaon, a blast-wave model is produced with a Monte Carlo simulation. Its temperature $T$ and transverse flow $\rho$ are extracted, and the comparisons of $T$ and $\rho$ between the fitting results and the setted values are shown. It is worth to note that the data of $m_\textmd T$ spectra and HBT radii $R_\text s(K_\text T)$ are fitted together, and a combined optimization function $\chi^2$ are deduced from a maximum likelihood method.  In Sec.~\ref{sec:ampt}, for A Multi-Phase Transport model (AMPT) \cite{ampt} at RHIC energy from 7.7 to 200 GeV, temperature $T$ and transverse flow rapidity $\rho$ are extracted by utilizing the shapes of both $m_\text T$ spectra and HBT radii $R_\text s(K_\text T)$, and the results are compared with the $T$ and $\rho$ extracted directly from AMPT model . A summary is given in Sec.~\ref{sec:summary}.

\section{Method\label{sec:method}}

\begin{figure}
\includegraphics[width=8.6cm]{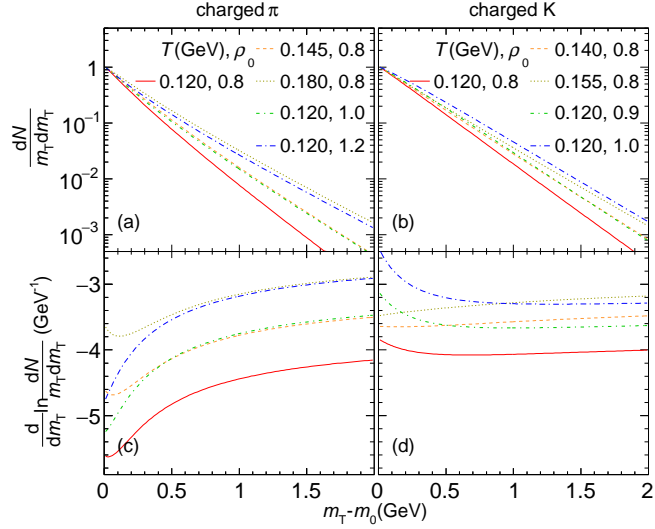}
\caption{Comparisons between $m_\text T$ spectra with different parameters $T$ and $\rho_0$ under the blast-wave function [Eqs.~(\ref{equ:BWE}) and (\ref{equ:mtSpectrum})] for charged pion and charged kaon are shown in upper pads. The corresponding graphs of $\frac{\mathrm d}{\mathrm d m_\text T } \ln \frac{\mathrm d N}{m_\text T \mathrm d m_\text T}$ [Eq.~(\ref{equ:dlnmts})] are shown in bottom pads for illustrating the shapes of $m_\text T$ spectra more clearly. \label{fig:mts}}
\end{figure}

In the blast-wave parametrization \cite{BlastWaveMtsRs2, BlastWaveMtsRs2_2}, it is assumed that the particles produced in $A$-$A$ collisions are local thermal equilibrium and freeze-out at a kinetic temperature $T$ with a transverse radial flow rapidity $\rho$ at the freeze-out surface. The emission function of only one kind of particle can be expressed as
\begin{equation}
S(x,p)\propto m_T \cosh\left(\eta-\mathrm y\right)\Theta\left(r\right)\mathrm e^{-\frac{(\tau-\tau_0)^2}{2\delta\tau^2}}
\mathrm e^{-\frac{p\cdot u}{T}},
\end{equation}
where, $x$ is the space-time of a particle, $p$ is 4-momentum of the particle, $\eta=\frac{1}{2} \ln {\frac{t+z}{t-z}}$ is called space-time rapidity, $\mathrm y=\frac{1}{2}\ln{\frac{E+p_z}{E-p_z}}$ is the rapidity, and $e^{-\frac{(\tau-\tau_0)^2}{2\delta\tau^2}}$ stands for the distribution of proper time. Transverse density distribution is described by $\Theta(r)$ which is assumed as a limited uniform distribution. Longitudinal expansion is boost invariant. Thermal distribution is described by local Boltzmann factor $e^{\frac{p \cdot u}{T}}$, and $u$ stands for a local expansion velocity. Transverse expansion rapidity increases with the distance to the center of the source linearly, so that if the transverse flow rapidity are denoted by $\rho_0$, the transverse flow rapidity for a particle with a radial distance $r$ can be expressed as $\rho = \rho_0 \frac{r}{R}$, where $R$ is the transverse radius of the limited uniform distribution \cite{BlastWaveMtsRs2, BlastWaveMtsRs2_2}. The emission function can be written as
\begin{equation}
\begin{aligned}
S\left(x,p\right)&\propto  m_T \cosh(\eta-\mathrm y)\Theta \left(r\right)
\mathrm e^{-\frac{\left(\tau-\tau_0\right)^2}{2\delta\tau^2}} \\
&\cdot \mathrm e ^{ \frac{p_T}{T}\sinh\rho\cos\left(\phi_p-\phi_s\right)
- \frac{m_T}{T}\cosh\rho\cosh\left(\eta-\mathrm y\right) }\\
\end{aligned}
\label{equ:BWE}
\end{equation}
The transverse mass spectrum can be expressed as\cite{BlastWaveMtsRs2, BlastWaveMtsRs2_2, LQShan, mts}
\begin{equation}
\begin{aligned}
&\frac{\mathrm dN}{m_\text T\mathrm dm_\text T} \propto \int S \left(x,p\right)\mathrm d \phi_p \mathrm d \mathrm y \mathrm d^4x   \\
&\!\!=\!\!A\!\!\int_0^R m_\text T I_0\left(\frac{p_\text T}{T}\sinh\rho\right)
K_1\left(\frac{m_\text T}{T} \cosh\rho  \right) r \mathrm dr\\
&\!\!=\!\!A'\!\!\int_0^1 m_\text T I_0\left(\frac{p_\text T}{T}\sinh\rho_0\xi\right)
K_1\left(\frac{m_\text T}{T} \cosh\rho_0\xi  \right) \xi \mathrm d\xi ,
\end{aligned}
\label{equ:mtSpectrum}
\end{equation}
where, $\phi_p$ is the azimuth of momentum, $A'=AR^2$ is a normalization factor, $I_0$ and $K_1$ are modified Bessel functions. In Eq.~({\ref{equ:mtSpectrum}}), radius parameter $R$ is absorbed by the normalization factor and does not influence the shape of $m_T$ spectrum, so the available parameters of the $m_T$ spectrum expression Eq.~(\ref{equ:mtSpectrum}) are $T$, $\rho_0$ and rest mass $m_0$ of the particles. The $m_\text T$ spectra for high $T$ and low $\rho_0$ is similar with the low $T$ and high $\rho_0$. For charged pion and charged kaon, this phenomenon is shown in Figs.~\ref{fig:mts}(a) and \ref{fig:mts}(b).

To see the shapes clearly and remove the confusion of normalization factor $A'$,  $\frac {\textmd d}{\textmd dm_T}\ln{\frac{\textmd dN}{m_T\textmd dm_T}}$ for different parameters $T$ and $\rho_0$ is shown in bottom  pads of Fig.~\ref{fig:mts} and can be expressed as\cite{mts}
\begin{equation}
\begin{aligned}
\frac {\mathrm d\ln{\frac{\mathrm dN}{m_\text T\mathrm dm_\text T}}}{\mathrm dm_\text T}= \frac{ \int_0^1{\left[ \frac{m_\text T^2}{p_\text T^2}\alpha I_1\left(\alpha\right)K_1\left(\beta\right)-\beta I_0\left(\alpha\right)K_0\left(\beta\right)  \right]\xi\mathrm d\xi} }
{  \int_0^1 m_\text T I_0\left(\alpha\right)K_1\left(\beta\right)\xi\mathrm d\xi },
\end{aligned}
\label{equ:dlnmts}
\end{equation}
where, $I_1\left(\alpha\right)$ and $K_0\left(\beta\right)$ are modified Bessel functions, $\alpha=\frac{p_T\sinh\rho}{T}$ and $\beta=\frac{m_T\cosh\rho}{T}$ are utilized to simplify the expression, and the expansion rapidities $\rho$ are equal to $\rho_0\xi$. Eq.~(\ref{equ:dlnmts}) reflects the shapes of transverse mass spectra. Fig.~\ref{fig:mts} shows that the $m_\text T$ spectra with increasing temperature parameter $T$ is similar the $m_\text T$ spectra with increasing transverse flow rapidity parameter $\rho_0$. This phenomenon suggest  a strong negative correlation of $T$ and $\rho_0$ for a measured $m_\text T$ spectrum. In other words, the $m_\text T$ spectrum can be changed similarly by rasing $T$ or rasing $\rho_0$.

On the other hand, the HBT radii $R_\text s \left(K_\text T\right)$ for identical pion or identical kaon are also sensitive to temperature $T$ and transverse flow rapidity $\rho_0$. The HBT radii $R_\text s$ decrease with the average transverse momentum $K_\text T$, and this phenomenon is considered to be dominated by the expansion of the source. Generally, the HBT correlation functions were considered to be Gaussian form in out-side-long reference system. For central collisions, because of the symmetry, HBT correlation function was expressed as \cite{hbt_sum, hbt_sum2, hbt_sum3, hbt_sum4}
\begin{equation}
C\left({\mathbf{q},\mathbf{K}}\right)=1+\lambda~\mathrm e^{-q_\text o^2 R_\text o^2\left({\mathbf K}\right)-q_\text s^2 R_\text s^2\left({\mathbf K}\right)-q_\text l^2R_\text l^2\left({\mathbf K}\right)},
\end{equation}
where the differences and averages of momentum of pairs of identical particles are expressed as $ \mathbf q = \mathbf p_1 - \mathbf p_2$ and $ \mathbf K=\frac{\mathbf p_1+\mathbf p_2}{2}$, respectively. Out-direction is along transverse projection of $\mathbf K$, long-direction is along the beam, side-direction is perpendicular to out-long plane, and $q_\text l$ is described in longitudinal comoving system(LCMS) of a pair of particles \cite{hbt_sum, hbt_sum2, hbt_sum3, hbt_sum4}. $R_\text s$ is independent of time parameter \cite{LQShan, hbt, hbt2, hbt3, hbt4, hbt_5, hbt_6, hbt_7, hbt_sum, hbt_sum2, hbt_sum3, hbt_sum4,  rs2, rs2_kt, rs2_rms}, and it can be expressed as
\begin{equation}
\begin{aligned}
R_s^2\left(\mathbf K \right)=\left\langle y^2 \right\rangle = \frac{\int{ y^2 S\left(x, \mathbf K\right)d^4 x}}{\int{S\left(x, \mathbf K\right)d^4 x}}.
\end{aligned}
\end{equation}
Under the assumption of blast-wave parametrization as Eq.~(\ref{equ:BWE}), $R_s^2$ can be expressed as \cite{hbt_sum, hbt_sum2, hbt_sum3, hbt_sum4, rs2_rms, rs2_kt, LQShan}
\begin{equation}
\begin{aligned}
R_s^2\left(K_\text T\right)&=\frac{ \int_0^R \frac{r^3}{\alpha} I_1\left(\alpha\right)K_1\left(\beta\right) \textmd dr }{ \int_0^R r I_0\left(\alpha\right)K_1\left(\beta\right) \textmd dr } \\
&= R^2 \frac{ \int_0^1 \frac{\xi^3}{\alpha} I_1\left(\alpha\right)K_1\left(\beta\right) \textmd d\xi } { \int_0^1 \xi I_0\left(\alpha\right)K_1\left(\beta\right) \textmd d\xi },
\end{aligned}
\label{equ:Rs2}
\end{equation}
where the equations $\alpha=\frac{K_\text T\sinh\rho}{T}$, $\beta=\frac{M_\text T\cosh\rho}{T}$, and $M_\text T = \sqrt{K_\text T^2 + m_0^2} $ are utilized for the simplification.

\begin{figure}
\includegraphics[width=8.6cm]{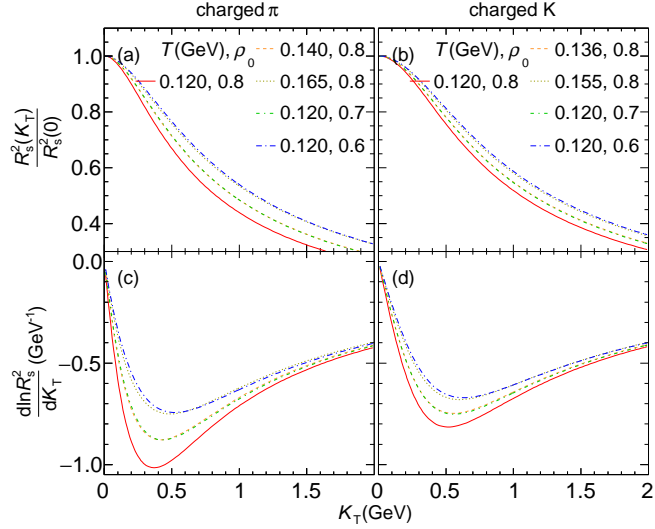}
\caption{HBT radii $R_\text s^2\left(K_\text T\right)$ [Eq.~(\ref{equ:Rs2})] are shown in upper pads for different parameters $T$ and $\rho_0$ under the blast-wave function [Eq.~\ref{equ:BWE}], and the $R_\text s^2\left(0\right)$ are utilized for the comparisons. The corresponding graphs of $\frac{\mathrm d \ln R_\text s^2}{\mathrm d K_\text T}$ [Eq.~(\ref{equ:dlnRs2})] are shown in bottom pads for illustrating the shapes of $R_\text s ^ 2 \left(K_\text T\right)$ more clearly. \label{fig:rs2}}
\end{figure}

If we aim at parameters $T$ and $\rho_0$, the shape of $R_s^2\left(K_\text T \right)$ can be reflected by $\frac{\mathrm d\ln R_\text s ^ 2}{\mathrm d K_\text T}$  expressed as
\begin{equation}
\begin{aligned}
&\frac{\mathrm d\ln R_s^{2}}{\mathrm d K_\text T}
= -\frac{ \int_0^1{ \left[ I_1(\alpha)K_1(\beta) - \frac{K_\text T^2\beta}{M_\text T^2}I_0(\alpha)\frac{K_0(\beta)+K_2(\beta)}{2} \right]\xi \mathrm d\xi} }
{ K_\text T\int_0^1{ I_0(\alpha)K_1(\beta)\xi \mathrm d\xi} }\\
&+\frac{ \int_0^1{\left[ I_0(\alpha)K_1(\beta)-I_1(\alpha)\frac{K_0(\beta) K_\text T ^2 \beta-K_1(\beta)(2M_\text T ^2 +K_\text T ^2)}{M_\text T ^2 \alpha} \right]\xi^3\mathrm d\xi} }
{ K_T\int_0^1{ I_1(\alpha)K_1(\beta)\frac{\xi^3}{\alpha} \mathrm d\xi } },\\
\end{aligned}
\label{equ:dlnRs2}
\end{equation}
where the parameter $R$ is removed, so that there are only two parameters $T$ and $\rho_0$ in Eq.~(\ref{equ:dlnRs2}). If we change the parameters by increasing $T$ or decreasing $\rho_0$, which are shown in Fig.~\ref{fig:rs2}, the similarity of the shapes of two ways means a strong positive correlation of $T$ and $\rho_0$. In other words, for $\frac{\mathrm d \ln R_\text s ^ 2}{\mathrm d K_\text T}$, the combination of high $T$ and high $\rho_0$ is equivalent to the combination of low $T$ and low $\rho_0$.

From the analysis above, it is possible to extract $T$ and $\rho_0$ by fitting $\frac{\mathrm d}{\mathrm d m_\text T }\ln\frac{\mathrm dN}{m_\text T \mathrm d m_\text T}$ and $\frac{\mathrm d \ln R_\text s ^ 2}{\mathrm d K_\text T}$ together. On the other hand, negative correlations of $T$ and $\rho_0$ from the shapes of $m_T$ spectra and the positive correlations of $T$ and $\rho_0$ from the shapes of $R_s\left(K_\text T\right)$ will be shown more clearly with equi-error contours for a Monte Carlo simulation of blast-wave function in next section.

\section{Monte Carlo simulation and data analysis\label{sec:mc}}

\begin{figure}
\includegraphics[width=8.6cm]{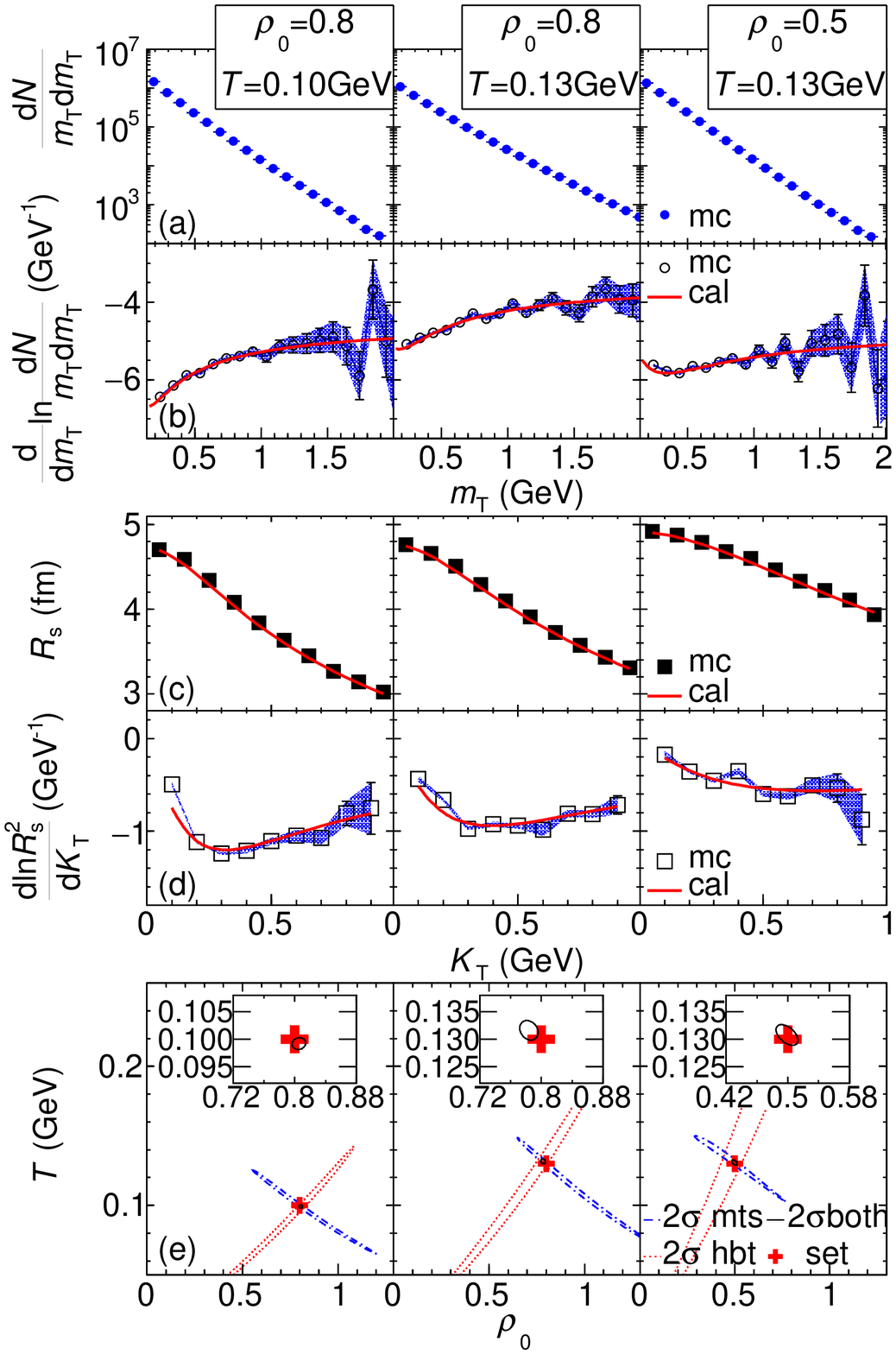}
\caption{The results of Monte Carlo simulation for charged pion. For different parameters $T$ and $\rho_0$, the particles are emitted by a Monte Carlo model under the blast-wave function [Eq.~(\ref{equ:BWE})]. The $m_\text T$ spectra, $\frac{\mathrm d}{\mathrm d m_\text T} \ln \frac{\mathrm d N}{m_\text T\mathrm d m_\text T}$, $R_\text T\left(K_\text T\right)$, and $\frac{\mathrm d \ln R_\text s^2}{\mathrm d K_\text T}$ of Monte Carlo model are shown in subgraphs (a), (b), (c), and (d), respectively, and the corresponding results of calculation with Eqs.~(\ref{equ:dlnmts}), (\ref{equ:Rs2}), and (\ref{equ:dlnRs2}) for the setted parameters are drawn by red curves. The 2$\sigma$ equi-error contours of fittings $\frac{\mathrm d}{\mathrm d m_\text T} \ln \frac{\mathrm d}{m_\text T\mathrm d m_\text T}$ ('2$\sigma$ mts'), $\frac{\mathrm d \ln R_\text s^2}{\mathrm d K_\text T}$ ('2$\sigma$hbt'), and both of them ('2$\sigma$ both') are shown in subgraphs (e), and the comparisons between setted $T$ and $\rho_0$ ('set') and results of combined fittings are enlarged in the inside pads. \label{fig:mc0}}
\end{figure}

\begin{figure}
\centering
\includegraphics[width=8.6cm]{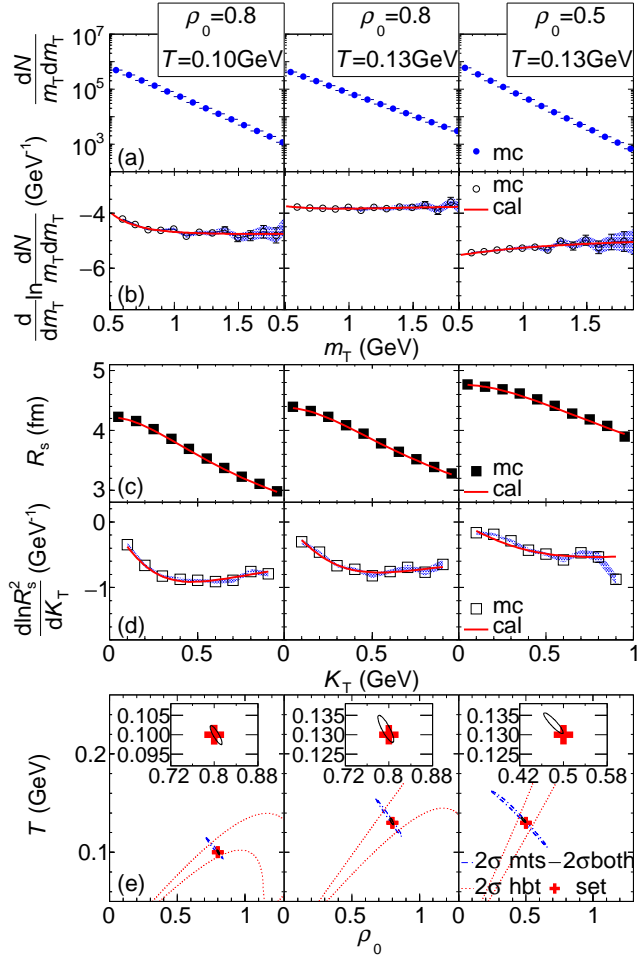}
\caption{The results of Monte Carlo simulation for charged kaon, and the explanations of the pads are same with Fig.~\ref{fig:mc0}. \label{fig:mc1}}
\end{figure}

For testing the method above, we simulate a Monte Carlo sources under the blast-wave function Eq.~(\ref{equ:BWE}) for identical pion and identical kaon respectively. The rest masses  $m_0$ of the particles are setted to mass of charged pion or charged kaon. Other parameters are set to $T=130\textmd{MeV}$, $\rho_0=\textmd{0.8}$, and $R=10\textmd{fm}$. We also change parameters $T$ and $\rho_0$ for testing this method, such as that $T$ decreases to $100$MeV or $\rho_0$ decreases to $0.5$. Two-particle correlations are calculated by the program Correlation After Burner(CRAB). The $m_\text T$ spectra, $\frac{\mathrm  d}{\mathrm dm_T}\ln{\frac{\mathrm dN}{m_\text T\mathrm dm_\text T}}$, $R_\text s\left(K_\text T\right)$, and $\frac{\mathrm d\ln R_\text s^{2}}{\mathrm d K_\text T}$ are shown in Figs.~\ref{fig:mc0} and \ref{fig:mc1} for identical pion and identical kaon respectively. On the $T$-$\rho_0$ plane [Figs.~\ref{fig:mc0}(e) and \ref{fig:mc1}(e)], the three kinds of contours stand for the results of fitting $\frac{\mathrm d}{\mathrm dm_\text T}\ln{\frac{\mathrm dN}{m_\text T\mathrm dm_\text T}}$ ('mts' in figures), $\frac{\mathrm d\ln R_\text s^{2}}{\mathrm d K_\text T}$('hbt' in figures), and both of them('both' in figures), respectively.

Before the discussions of the results, it may be needed to introduce the basic principle of the fitting simply, because two kinds of data ($m_\text T$ spectra and HBT radii $R_\text s$) are utilized together in the fitting. Fittings in this paper are made by utilizing least square method which is a special case of maximum likelihood method. It is assumed that the errors obey Gaussian distributions, and the values of errors stand for half widths of the distributions. For a transverse mass spectrum,  we denote the expression $\frac{\mathrm d}{\mathrm d m_\text T} \ln \frac{\mathrm d N}{m_\text T \mathrm d m_\text T}$ [Eq.~(\ref{equ:dlnmts})] to $f\left(m_\text T; T, \rho_0\right)$ where $m_\text T$ is a variable, $T$ and $\rho_0$ are parameters. For a point of measured $\frac{\mathrm d}{\mathrm d m_\text T}\ln\frac{\mathrm d N}{m_\text T \mathrm d m_\text T}$, which is denoted as ($m_\text {Ti}$, $f_\text i\pm\sigma_{f\text i}$), the probability can be expressed as
\begin{equation}
P_{f\text i}=\frac{1}{\sigma_{f\text i}\sqrt{2\pi}}
\exp\left\{-\frac{\left[f_\text{i}-f\left(m_\text{Ti}; T, \rho_0\right)\right]^2}{2\sigma_{f\text i}^2}\right\}.
\end{equation}
For a number of measured $(m_\text{Ti}, f_\text i\pm\sigma_{f\text i})$ points (i = 1, 2, ..., $n$), the probability $P_f$ of the data group can be expressed as the product of $P_{f\text i}$
\begin{equation}
\begin{aligned}
P_f=\prod_{i}{P_{f\text i}}={\frac{\left(2\pi\right)^{-\frac{n}{2}}}{\prod_{i}\sigma_{f\text i }}}\mathrm e^{-\frac{1}{2}\sum_\text i \frac{\left[f_\text{i}-f\left(m_\text{Ti}; T, \rho_0\right)\right]^2} {\sigma_{f\text i}^2}}
\label{equ:PY}
\end{aligned}
\end{equation}
where $n$ is the number of the points. The factor $\prod_{i}{\left(\frac{1}{\sigma_i\sqrt{2\pi}}\right)}$ is a constant for the measured points, so when the maximum of $P_f$ is needed, the summation $\chi_{f}^2\equiv \sum_\text i  \frac{1}{\sigma_{f\text i}^2} \left[f_\text{i}-f\left(m_\text{Ti}; T, \rho_0\right)\right]^2$ should be minimized.


It is similar for $\frac{\mathrm d \ln R_\text s ^2 }{\mathrm d K_\text T}$, which describes the shape of HBT radius $R_\text s ^ 2(K_\text T)$, and Eq.~(\ref{equ:dlnRs2}) are denoted to $h(K_\text T; T, \rho_0)$. The probability for some measured points $\left(K_\text{Tj}, h_\text j\pm\sigma_\text{hj}\right)$ of $\frac{\mathrm d \ln R_\text s ^2 }{\mathrm d K_\text T}$ (j = 1, 2, ..., $m$) can be expressed as
\begin{equation}
P_h=\frac{\left(2\pi\right)^{-\frac{m}{2}}}{\prod_\text j\sigma_{h\text j}^2} ~\mathrm e^{ -\frac{1}{2}\sum_{\text j}\frac{\left[h_\text j - h\left(K_\text{Tj};T,\rho_0\right)\right]^2}{\sigma_{h\text j}^2}},
\end{equation}
where $m$ is the number of the points, and when the maximum of $P_h$ is needed, the minimization of $\chi_{h}^2\equiv \sum_\text j \frac{1}{\sigma_{h\text j}^2} \left[h_\text j -h\left(K_\text {Tj}; T, \rho_0\right)\right]^2$ should be made.

When two kinds of data are both used, for the optimization of parameter $T$ and $\rho_0$, the probability $P_{fh}$ of observing these two group of points should be maximized, and $P_{fh}$ can be expressed as the product of $P_f$ and $P_h$,
\begin{equation}
\begin{aligned}
P_{fh}&=P_{f}\cdot P_{h}\\
&=a~\mathrm e^{
-\sum_\text i \frac{\left[f_\text{i}-f\left(m_\text{Ti}; T, \rho_0\right)\right]^2} {2\sigma_{f\text i}^2}
-\sum_{\text j}\frac{\left[h_\text j - h\left(K_\text{Tj};T,\rho_0\right)\right]^2} {2\sigma_{f\text j}^2} }\\
&=a~\mathrm e^{-\frac{1}{2}\left(\chi_f^2+\chi_h^2\right)},
\end{aligned}
\end{equation}
where $a=\prod_\text i \frac{1}{\sigma_{f\text i} \sqrt{2\pi}}
\prod_{\text j}\frac{1}{\sigma_{h\text j}\sqrt{2\pi}}$ is a constant for the measured points. When the optimization of the parameters $T$ and $\rho_0$ is needed, $P_{fh}$ should be maximized so that $\chi_f^2+\chi_h^2$ should be minimized.

Results of fitting are shown in Figs.~\ref{fig:mc0} and \ref{fig:mc1}. The setted ($T$, $\rho_0$) points are marked on $T$-$\rho_0$ plane as red cross. To see the results clearly, the contours for the combined fitting are enlarged on inside pads of Figs.~\ref{fig:mc0}(e) and \ref{fig:mc1}(e).  From the equi-error contours of fitting $\frac{\mathrm d}{\mathrm dm_\text T}\ln\frac{\mathrm dN}{m_\text T\mathrm dm_\text T}$, the negative correlations of $T$ and $\rho_0$ are shown clearly. The positive correlations of $T$ and $\rho_0$ from HBT results are also obvious for charged pion. If the surface transverse flow rapidity $\rho_0$ is small, the positive correlation of $T$ and $\rho_0$ is obvious for charged kaon, such as $T$ and $\rho_0$ are set to 0.130 GeV and 0.5, respectively. When $\rho_0$ is big, the positive correlations are indistinct for charged kaon such as $\rho_0$ is set to 0.8, but this phenomenon do not affect the results of fitting two kinds of data together. It is notable that the contours of combined fitting are near the cross, and the setted value of parameters are gotten.  In conclusion, the setted parameters are extracted successfully from the Monte Carlo model. In next part, this method will be used for extracting $T$ and $\rho_0$ from AMPT model of Au+Au collisions at $\sqrt{s_\text{NN}}$ = 7.7 to 200 GeV, and the results will be compared with the corresponding parameters extracted directly from the model.

\section{Extracting temperature and transverse flow from AMPT model\label{sec:ampt}}

\begin{figure}
\centering
\includegraphics[width=8.6cm]{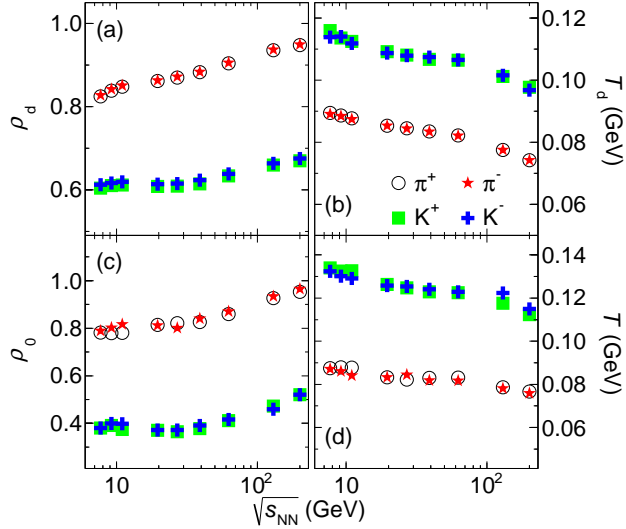}
\caption{The results of $T$ and $\rho$ for Au+Au collisions AMPT model at $\sqrt{s_\text{NN}}$ from 7.7 to 200 GeV. The results of combined fittings and direct extractions are shown in bottom and upper pads, respectively.\label{fig:ampt}}
\end{figure}

In this section, the combined fitting method is used for A Multi-Phase Transport model (AMPT) \cite{ampt}. AMPT model is successful for $m_\text T$ spectra of hadrons, HBT correlations and some other observations at RHIC energy. For AMPT of Au+Au collisions at RHIC energy $\sqrt{s_\text {NN}}$ = 7.7 to 200 GeV, the $T$ and $\rho_0$ are extracted by fitting $\frac{\mathrm d}{\mathrm d m_\text T} \ln \frac{\mathrm dN}{m_\text T \mathrm d m_\text T}$ and $\frac{\mathrm d \ln R_\text s ^2}{\mathrm d K_\text T}$ together, and the results are shown in the bottom pads of Fig.~\ref{fig:ampt}. For checking the fitting results, the temperature and transverse flow rapidity are also extracted from AMPT model directly, and they are denote to $T_\text d$ and $\rho_d$ respectively.

$T_\text d$ and $\rho_d$ are calculated as following steps.
First, the AMPT model is ran as central collisions, and the events of same collision energy compose a big event in our research, so that the particles are chosen from the big event ignoring the native events. Second, taking $\pi^-$ as an example, 3-momentum $\mathbf{p}$ of a particle in the centre-of-mass frame of a source is decomposed into three parts ($p_r$, $p_\phi$, $p_l$), where $p_r$, $p_\phi$, and $p_l$ are the projections in radial, tangential, and longitudinal directions, respectively. Because $p_\phi$ is not influenced by the transverse and longitudinal flow for central collisions, $p_\phi$ can be used for extracting $T_\text d$. At the intermediate rapidity range ($\left|\eta\right| < 0.5$), the spectra of $p_\phi$ can be expressed as
\begin{equation}\label{equ:p_phi_spectrum}
\frac{\mathrm d N}{\mathrm d p_\phi} \propto \int_{-\infty}^{\infty}\mathrm e^{-\frac{1}{T_\text d}\sqrt{p_\phi^2 +{p'}_r^2+ m_0^2}}\mathrm d {p'}_r,
\end{equation}
where ${p'}_r$ can be considered to be the radial momentum of a particle in the centre-of-mass frame of a cell, and the equation is deduced from a Boltzmann distribution. $T_\text d$ can be extracted by fitting $p_\phi$ spectra with Eq.~(\ref{equ:p_phi_spectrum}). At last, the transverse flow rapidities of a source can be estimated by the averages of the radial rapidities of the cells which compose the source. The radial velocity  of a cell can be expressed as $\beta_\text{cell} = \sum p_r / \sum E$,
where $E$ is the energy of the chosen particle in a cell. The radial rapidity $\rho_\text{cell}$ of the cell can be estimated as $\rho_\text {cell} = \frac{1}{2}\ln\left[\left(1+\beta_\text {cell}\right)/\left(1-\beta_\text {cell}\right)\right]$.
The average transverse flow rapidity are expressed as $\rho_\text d = \langle \rho_\text{cell}\rangle$, and the normalization are made by utilizing the numbers of the chosen particles in the cells.

For $\pi^+$, $\pi^-$, $K^+$, and $K^-$, the results of AMPT model are shown in Fig.~\ref{fig:ampt}. The results of $T$ and $\rho_0$ for combined fitting are shown in bottom pads, and $T_\text d$ and $\rho_\text d$ from the direct extraction are shown in upper pads.  $T_\text d$ decrease with the $\sqrt{s_\text{NN}}$ obviously, and $\rho_\text d$ are all increasing with the collision energy $\sqrt{s_\text {NN}}$. The tendencies of the fitted $T$ and $\rho_0$ vs $\sqrt{s_\text {NN}}$ are similar with the parameters $T_\text d$ and $\rho_\text d$ extracted directly from the model, respectively. In the other words, the tendencies of $T_\text d$ and $\rho_\text d$ vs $\sqrt{s_\text {NN}}$ are reproduced by the combined fitting. On the other hand, the values of $T_\text d$ of pion are lower obviously than those of kaon, and the transverse flow rapidities $\rho_\text d$ of pion are higher obviously than those of kaon. These phenomena are also observed for $T$ and $\rho_0$ using the combined fitting. We guess that heavier meson may freeze out earlier and not cool down enough, so that the time of being pushed by the transverse flow may be shorter for kaon.

\section{Summary\label{sec:summary}}

Based on a blast-wave function, we analyze the shapes of single-particle transverse mass $m_\text T$ spectra and HBT radius $R_\text s(K_\text T)$ for identical pion and identical kaon. The negative correlations of thermal temperature $T$ and surface transverse flow rapidity $\rho_0$ are shown by $\frac{\textmd d}{\textmd d m_T}\ln\frac{\textmd dN}{m_T\textmd dm_T}$, and positive correlations of $T$ and $\rho_0$ are also shown clearly by $\frac{\textmd d\ln R_s^2}{\textmd dK_T}$. Combined fitting of $\frac{\textmd d}{\textmd d m_T}\ln\frac{\textmd dN}{m_T\textmd dm_T}$ and $\frac{\textmd d\ln R_s^2}{\textmd dK_T}$ shows a way to get temperature and radial flow rapidity utilizing a single kind of meson (identical pion or identical kaon). With a Monte Carlo model, the equi-error contours shows that the combined fitting method works well. At last, this method is used for AMPT model of Au+Au collisions at RHIC energy for  $\sqrt{s_\text{NN}}$ = 7.7 to 200 GeV. The results of combined fitting method shows same tendencies of $T\left(\sqrt{s_\text {NN}}\right)$ and $\rho\left(\sqrt{s_\text {NN}}\right)$ with the corresponding parameters $T_\text d\left(\sqrt{s_\text {NN}}\right)$ and $\rho_\text d\left(\sqrt{s_\text {NN}}\right)$ extracted directly from the model, respectively. For both of the methods (combined fitting and direct extraction), the parameters temperature $T$ and transverse flow rapidity $\rho$ of identical pion are both different from those of identical kaon. This phenomenon suggests that it is necessary to extract temperature and transverse flow parameters with a single kind of particle.

\begin{acknowledgments}
We thank Wei-Ning Zhang and Yan Yang for their advices and discussions.
\end{acknowledgments}

%
%


\end{document}